# Urban Mobility in the Age of Automation: Analyzing Public Attitudes Toward Privately-Owned versus Shared Automated Vehicles

Yellitza Soto, Fatemeh Nazari[1], Mohamadhossein Noruzoliaee

*Department of Civil Engineering, The University of Texas Rio Grande Valley, Edinburg, TX 78539, United States*

**Abstract**: Transportation systems will be likely transformed by the emergence of automated vehicles (AVs) promising for safe, convenient, and efficient mobility, especially if used in shared systems (shared AV or SAV). However, the potential tendency is observed towards owning AV as a private asset rather than using SAV. This calls for research on investigating individuals' attitude towards AV in comparison with SAV to recognize the barriers to the public's tendency towards SAV. To do so, the present study proposes a modeling framework based on the theories in *behavioral psychology* to explain individuals' preference for owning AV over using SAV, built as a *latent (subjective) psychometric construct*, by three groups of explanatory latent constructs including: (i) desire for searching for benefits, i.e., extrinsic motive manifested in *utilitarian beliefs*; (ii) tendency towards seeking pleasure and joy, i.e., intrinsic motive reflected in *hedonic beliefs*; and (iii) attitude towards three configurations of shared mobility, i.e., experience with car and ridesharing, bikesharing, and public transit. Estimated on a sample dataset from the State of California, the findings can shed initial lights on the psychological determinants of the public's attitude towards owning AV versus using SAV, which can furthermore provide policy implications intriguing for policy makers and stakeholders. Of note, the findings reveal the strongest influential factor on preference for AV over SAV as hedonic beliefs reflected in perceived enjoyment. This preference is next affected by utilitarian beliefs, particularly perceived benefit and trust of stranger, followed by attitude towards car and ride sharing.

[1] Corresponding author: Fatemeh Nazari, fatemeh.nazari@utrgv.edu

## 1. Introduction

The future of transportation systems is evolving with the emergence of two technological advances — i.e., automated vehicles (AVs) and shared mobility. AVs are capable of operating without any human interaction, and thus can provide safer mobility, and less traffic congestion and fuel emissions (Fagnant and Kockelman, 2015; Andreson et al., 2016). Shared mobility, on the other hand, is gaining recognition with its popular on-demand services such as carsharing, ridesharing, and ride-hailing which are cost-efficient due to transporting multiple individuals at once (Shaheen et al., 2016). The conflation of these two trends leads to the idea of shared AVs (SAVs) promising to exploit benefits of both (Lee et al., 2019; Narayanan et al., 2020). Notwithstanding the SAV benefits, there are still a potential demand for private ownership of AVs evidenced by the literature findings (Haboucha et al., 2017; Lavieri et al., 2017; Nazari et al., 2018). While the existing studies provide valuable insights, there is a research gap of recognizing factors influencing the public's tendency towards privately owning AV in comparison with using SAV.

To fill that gap, the present research explores behavioral preferences to propose a modeling framework based on *behavioral psychology*. In the proposed model, preference for AV over SAV, which is built as a *latent (subjective) psychometric construct*, is explained by three groups of explanatory latent constructs including: (i) desire for searching for benefits, i.e., extrinsic motive manifested in *utilitarian beliefs*; (ii) tendency towards seeking pleasure and joy, i.e., intrinsic motive reflected in *hedonic beliefs*; and (iii) attitude towards three configurations of shared mobility, i.e., car and ridesharing, bikesharing, and public transit. Estimated on a sample dataset from the State of California (2019), the study findings can shed initial lights on the psychological determinants of the public's attitude towards owning AV and using SAV, which can furthermore provide policy implications intriguing for policy makers and stakeholders.

The structure of the paper follows with the next section discussing the conceptual framework and the study hypothesis. Next, a section is on the statistical analysis of the dataset used for the empirical estimation and another section on discussing the empirical estimation results. The last section concludes with a summary of the main findings.

## 2. Conceptual Framework and Hypotheses

A large body of the existing literature with the *psychological* perspective explores acceptance of technology by connecting individuals' acceptance behavior to their traits, attitudes, and preferences. In the context of (S)AV acceptance and adoption behavior, a stream of studies apply the theory of technology acceptance model (Davis et al., 1989) to explain individuals' behavioral intention to use (S)AVs by two core latent constructs describing perceived usefulness and perceived ease of use (Lee et al., 2019; Panagiotopoulos and Dimitrakopoulos, 2018; Zhang et al., 2019). Built on the theory of planned behavior (Ajzen, 1991), another stream of studies explains individuals' attitude towards (S)AV by attitudes, subjective norms, and behavioral control (Chen and Chao, 2011). Another approach is the unified theory of acceptance and use of technology (Venkatesh et al., 2003) consisting of four predictors (i.e., performance expectancy, effort expectancy, social influence, and facilitating conditions) as the foundation for users' behavioral intentions towards (S)AV (Madigan et al., 2017).

Inspired by the above-discussed theories, the present research proposes a modeling framework which is shown in Figure 1. The dependent variable is a latent construct explaining individuals' preference for owning AV over using SAV built on an observed item. The explanatory variables are latent constructs categorized in three groups including utilitarian beliefs, hedonic beliefs, and attitude towards shared mobility. *Utilitarian beliefs*, which relates to individuals' rational decision making, is indicated by three latent constructs including perceived benefit, perceived safety of other road users than the AV passengers, and trust of strangers, which are built on four, one, and one observed items, respectively. *Hedonic beliefs*, which mostly expresses emotional decision making, is captured by one latent construct explaining perceived enjoyment built on two items. Lastly, individuals' *attitude toward shared mobility* is addressed by three latent constructs explaining their experience with car and ride sharing, bikesharing, and public transit which are respectively built on five, two, and three items. The eight latent constructs and the corresponding observed items are listed in Table 1.

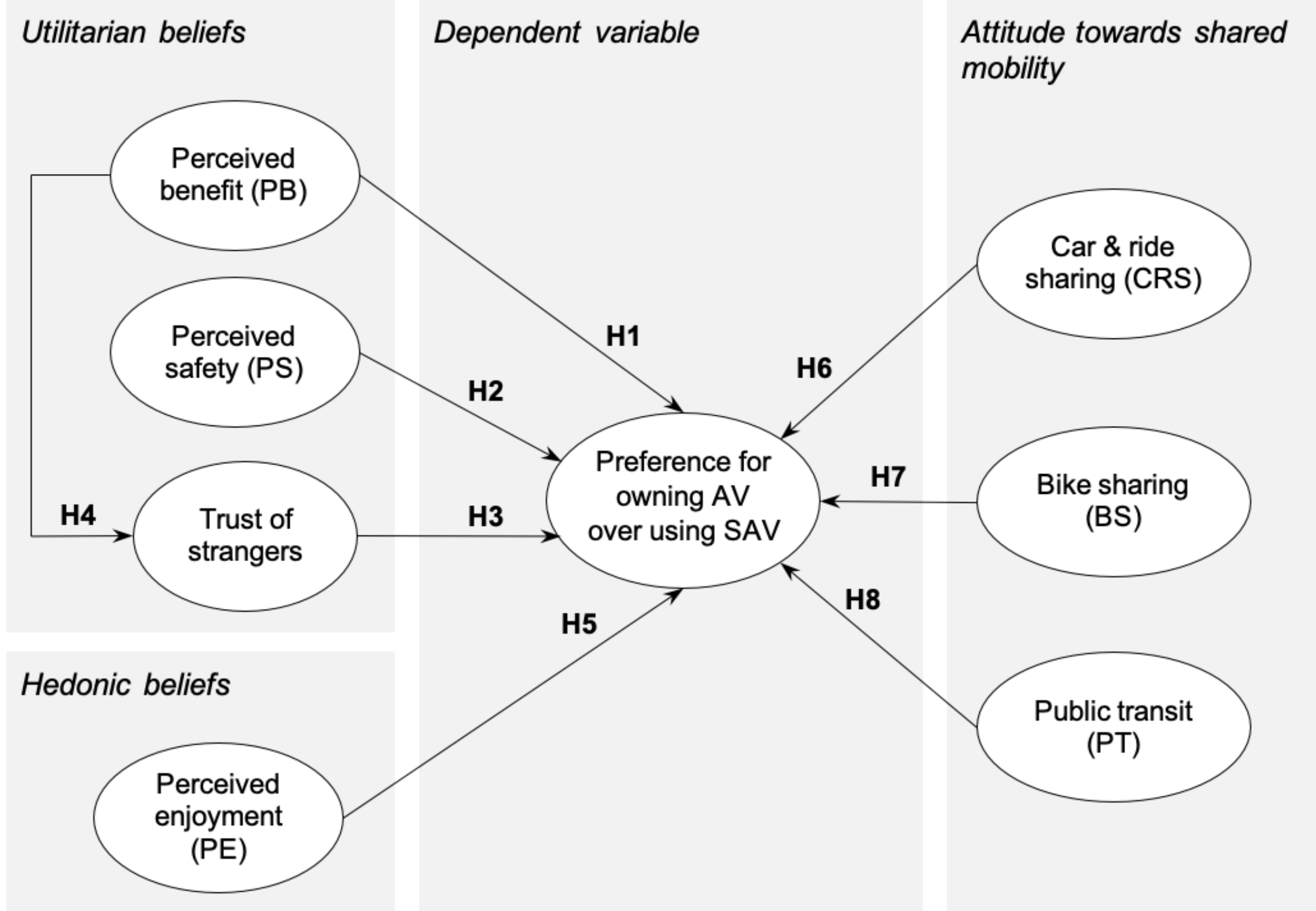

**Figure 1: The proposed model of preference for owning AV over using SAV**

The proposed framework is estimated using structural equation modeling (SEM) technique including two sets of measurement and structural equations, where the former yields the weight of each item on the corresponding latent construct. The latter determines the potential impacts of the latent constructs on each other defined as hypotheses, as shown in Figure 1. The remainder of this section justifies the definition of the eight latent constructs and the associated items along with discussing the eight hypotheses.

### 2.1. Dependent variable: Preference for owning AV over using SAV

Referring to the discussion in the introduction section, there exists a research need for better understanding the psychological factors affecting individuals' preference for having an AV as an owned asset versus using an AV provided in the shared transportation system. Based on that, the outcome variable of this study is the individuals' behavioral preference for owned versus shared AVs.

**Table 1. Latent constructs and the corresponding items**

| Latent construct | Definition |
|---|---|
| ***Dependent variable*** | |
| Preference for owning AV over using SAV (AV over SAV) | |
| AV over SAV | Overall, what would be your relative interest in owning a driverless vehicle versus using on-demand ride-hailing services? |
| ***Utilitarian beliefs*** | |
| Perceived benefit (PB) | |
| PB1 | I would reduce my time at the regular workplace and work more in the self-driving car. |
| PB2 | I would send an empty self-driving car to pick up/drop off my child. |
| PB3 | I would be able to travel more often even when I am tired, sleepy, or under the influence of alcohol/medications. |
| PB4 | I see a need for self-driving vehicles. |
| Perceived safety of other road users than the AV passengers (PS) | |
| PS1 | I would accept longer travel times so the self-driving vehicle could drive at a speed low enough to prevent unsafe situations for pedestrians and bicyclists. |
| Trust of strangers | |
| Trust1 | I would be likely to use shared driverless services (even at higher cost) because I would have no issues with sharing a vehicle with strangers. |
| ***Hedonic beliefs*** | |
| Perceived enjoyment (PE) | |
| PE1 | A self-driving vehicle would enable me to enjoy traveling more (e.g., watch scenery, rest). |
| PE2 | I would not miss the joy of driving and being in control. |

| Latent construct | Definition |
|---|---|
| ***Attitude towards shared mobility*** | |
| Car and ride sharing experience (CRS) | What is your experience with the following transportation options for trips in your local area? |
| CRS1 | Rental car |
| CRS2 | Ride-hailing (Uber/Lyft) |
| CRS3 | Shared ride-hailing (UberPool/LyftLine) |
| CRS4 | Carsharing (Car2Go, ZipCar) |
| CRS5 | Peer-to-peer car rental (e.g., GetAround, Turo) |
| Bike sharing experience (BS) | What is your experience with the following transportation options for trips in your local area? |
| BS1 | Bikesharing (e.g., Bay Area Bike Share) |
| BS2 | Shared eBikes or eScooters (e.g., Jump) |
| Public transit experience (PT) | What is your experience with the following transportation options for trips in your local area? |
| PT1 | Public bus |
| PT2 | Light rail/tram/subway (e.g., BART, LA Metro) |
| PT3 | Commuter train (e.g. Amtrak, Caltrain) |

### 2.2. Antecedent variables: Utilitarian and hedonic beliefs

Motivation energizes or activates a person to act towards achieving a goal. According to the definition proposed in the psychological self-determination theory (Deci and Ryan, 2013), motivations have different types rooted in the primary reason giving rise to the action, among them, two fundamental are *intrinsic* and *extrinsic* motivations. The former refers to "doing something because it is inherently interesting or enjoyable", whereas the latter is defined as the "doing something because it leads to a separable outcome" (Ryan and Deci, 2000). In the motivational technology acceptance model (Davis et al., 1992), the behavioral intention to use a technology for *intrinsically motivated* persons is because they realize it "for no apparent reinforcement other than the process of performing the activity per se", whereas for *extrinsically motivated* persons is because they perceive the technology as "instrumental in achieving valued outcomes that are distinct from the activity itself, such as improved job performance, pay, or promotions."

On the comparison of the two motivation types, a meta-analytic study (Gerow et al., 2013) along with many empirical studies exploring the user acceptance behavior of a system find that intrinsic motivation (e.g., fun, fear, sensory stimulation, and joy) is a stronger determinant of *hedonic* application use of the system (Van der Heijden, 2004), and extrinsic motivation (e.g., usefulness, benefit, and performance expectancy) is superior to determine *utilitarian* application use of the system (Venkatesh et al., 2003). In light of this, intrinsic and hedonic, and extrinsic and utilitarian are used interchangeably through the literature. There is a large body of the literature investigating the influence of hedonic factors, motivating a person's behavior as the results of stimulating pleasure or pain receptors, and utilitarian factors, motivating a person's behavior driven by satisfying basic needs and solving problems, on behavioral intention to use/accept/adopt AVs (Herrenkind et al., 2019; Keszey, 2020; Kapser and Abdelrahman, 2020).

There exists a growing body of the literature exploring whether and how individuals' behavioral intention towards AVs are affected by utilitarian factors such as perceived usefulness/benefit and performance expectancy (Madigan et al., 2017) and hedonic factors such as perceived enjoyment, according to a review study (Keszey, 2020). However, an understanding of the role of these factors in preference for owning an AV versus using AV in a shared mobility system is lacking. Thus, two groups of constructs are defined including utilitarian and hedonic beliefs, as shown in the left side of Figure 1. The former includes three latent constructs explaining perceived benefit, perceived safety and, and trust. The latter captures individuals' joy through the perceived enjoyment latent construct.

*2.2.1. Perceived benefit*

Prior studies find that individuals likely accept and adopt a technology if they perceive it to be beneficial or useful (Siegrist, 2000; Terwel et al., 2009). In the motivational technology acceptance model (Davis et al., 1989), perceived usefulness is a core factor directly influencing the individuals' behavioral intention to use the technology. In the context of AV technology, understanding the perceived benefit is the objective of a number of studies comprehensively reviewed in Gkartzonikas and Gkritza (2019).

There is a number of relevant studies finding perceived usefulness/benefit as a primary factor on behavioral intention to use AV, according to a relevant review study (Keszey, 2020). For instance, Ward et al. (2017) find this factor to have stronger role in AV acceptance than prior knowledge, perceived trust, and perceived risk. In another example, perceived benefit is found to be a stronger predictor of AV acceptance than perceived risk (Liu et al., 2019b). This factor is also found to increase positive attitude towards behavioral intention to user (Choi and Ji, 2015; Zhang et al., 2019), willingness-to-use (Hohenberger et al., 2017; Liu et al., 2019a), and public attitude towards AV (Liu, 2020). Based on these observations, the first hypothesis of the present study is defined as below.

**Hypothesis 1 (H1):** Perceived benefit positively influences preference of owning AV over using SAV.

*2.2.2. Perceived safety of other road users than the AV passengers (PS)*

The results of prior surveys and studies reveal that the respondents express their concern about safety aspects of AVs. For instance, a survey of 467 individuals reveals that the most influential factor on attitude toward AVs is stated as safety and legal/regulatory issues respectively by 82% and 12% of the respondents (Casley et al., 2013). Another survey asks the opinion of 5,000 respondents who express that their most concern is about software hacking/misuse, legal issues, and safety (Kyriakidis et al., 2015). The respondents' concern is confirmed to play a major role in AV acceptance/adoption behavior, according to the findings of the empirical studies (Xu et al., 2018; Dannemiller et al., 2021), a review study (Keszey, 2020) confirming the significant impact of individuals' perceived safety on their behavioral intention to use AVs, and another review study (Gkartzonikas and Gkritza, 2019) mentioning the concern about safety issues as barriers to AV deployment.

In the existing studies however, two gaps are observed which are aimed to be addressed by the present study. First, the safety concern factor mainly relates to the (potential) users of the AV system, while AVs will ride on roads interacting with the environment encompassing other road users such as pedestrians and bicyclists. Therefore, a person's tendency towards the AV technology needs to be explored considering his/her concern about the safety of other road users than the AV passengers (Ackermann et al., 2019; Combs et al., 2019; Deb et al., 2017; Velasco et al., 2019). Second, the majority of the related studies focus only on the privately-owned AVs, paying less attention to the individuals' tendency to SAV against private AVs. A relevant work is an empirical study (Nazari et al., 2018) assessing how individuals' interest in privately-owned AVs as well as four SAV configurations is affected by their perceived safety concern. Their findings reveal the dominant role of the safety factor on both AVs and the four SAV configurations. To fill these two gaps, the second hypothesis is formed as below.

**Hypothesis 2 (H2):** Perceived safety of other road users than the AV passengers positively influences preference of owning AV over using SAV.

*2.2.3. Trust*

The prior studies agree that a person's attitude towards a particular technology is significantly affected by his/her trust in the technology which is defined as the person's comfort in placing him or herself in an uncertain and vulnerable situation with respect to the technology expecting to gain benefit or utility (Mayer et al., 1995). Examples of this finding are in the acceptance and adoption of automated technology (Lee and Moray, 1992; Verberne et al., 2012) and the AV technology (Shariff et al., 2017; Xu et al., 2018; Zhang et al., 2019; Hegner et al., 2019). The relevant empirical studies focusing on the AV technology are thoroughly reviewed in Keszey (2020).

In the context of accepting/adopting to use AV in a shared system (i.e., SAV where an individual may share an AV with other passengers), in addition to trust in the technology, the individual's trust in sharing the AV with strangers needs to be considered. For instance, to explore the attitude towards ridesharing services, Bansal et al. (2016) look into the associated role of comfort in sharing a ride with three types of persons including strangers for a short duration in daytime, Facebook friends (never met before), and friends or family members. In an empirical study on user's preferences for AVs versus SAVs, Haboucha et al. (2017) investigate the related impact of public transport attitude defined as a latent construct which includes an item on individuals' comfort in riding public transit with strangers.

Looking at the literature on the acceptance and adoption of SAV, especially in comparison with owning AVs, reveals the rare attention paid to the related role of individuals' willingness to share a ride with strangers. In response, the present study hypothesizes the negative influence of individuals' trust of strangers in their tendency towards owning AV versus using SAV. Moreover, according to the definition discussed above, individuals' trust in a system is to the end of the desirable outcome or the pursuit of benefit. Accordingly, the fourth hypothesis of this study is assessing the positive impact of perceived benefit on trust of strangers.

**Hypothesis 3 (H3):** Trust in sharing a vehicle with strangers positively influences preference of owning AV over using SAV.

**Hypothesis 4 (H4):** Trust in sharing a vehicle with strangers is positively influenced by perceived benefit.

*2.2.4. Perceived enjoyment*

Recalling the discussion at the beginning of the section, individuals' tendency towards a particular system not only is the result of searching for benefits, i.e., extrinsic motive manifested in utilitarian beliefs, but due to seeking pleasure and joy, i.e., intrinsic motive reflected in hedonic beliefs. To verify, hedonic beliefs are reported to strongly affect the technology acceptance behavior is the various fields (Van der Heijden, 2004). In the area of travel behavior, car use is reported to be accompanied with individuals' natural enjoyment in the act of driving or having control of the wheel due to forthcoming pleasure. For instance, Steg (2005) finds that car use is influenced by hedonic beliefs represented by affective motive which is defined by statements such as "I love driving" and "I like to drive just for the fun."

Hence, it is expected to observe driving enthusiasts resistant to accept an AV requiring the passengers to delegate the driving task to the AV. This expected observation is reported for AV which is found to be more favorable to individuals who positively respond to the statement " I'm just not interested in cars" than those who are driving enthusiasts revealing enjoyment for driving (Ipsos MORI, 2014). In another early study, Silberg (2013) relates individuals' passion for driving to their interest in AVs which results in a finding that early AV enthusiasts are among those who have low passion for driving. Later on, through the investigation of individuals' choice among conventional car, privately-owned AV, and SAV, Haboucha et al. (2017) reports that individuals' who enjoy driving likely prefer using their regular car over an AV or SAV.

In the recent research attempts, a number of empirical studies detect the negative impact of the enjoyment of driving a car as a factor negatively impacting AV acceptance (Hegner et al., 2019) and adoption (Asgari and Jin, 2019), while a few other studies report hedonic motivation built on enjoyment and pleasure from using automated technology positively affecting tendency towards automated road transport systems (Madigan et al., 2017), SAV use (Yuen et al., 2020), and attitude towards and behavioral intention to use autonomous shuttle (Chen, 2019). While providing valuable insights, there is still a research need for better understanding the role of perceived enjoyment which is a hedonic belief on the individuals' attitude towards AVs, especially in the comparison of privately-owned AVs and SAVs. For this purpose, the fifth hypothesis of the present study is defined as below.

**Hypothesis 5 (H5):** Perceived enjoyment positively influences preference of owning AV over using SAV.

*2.2.5. Attitude towards shared mobility*

Personal lifestyle such as travel attitude is found to affect individuals' propensity for AV and SAV acceptance and adoption by empirical studies (Chen and Chao, 2011; Lavieri et al., 2017; Nazari et al., 2018) and a relevant review study (Becker and Axhausen, 2017). Given the outcome variable of this study, we focus on the individuals' current travel attitude towards shared mobility captured by three latent constructs explaining car and ride sharing, bikesharing, and public transit experience. Accordingly, the last three study hypotheses are defined as below.

**Hypothesis 6 (H6):** Car and ride sharing experience positively influences preference of owning AV over using SAV.

**Hypothesis 7 (H7):** Bike sharing experience positively influences preference of owning AV over using SAV.

**Hypothesis 8 (H8):** Public transit experience positively influences preference of owning AV over using SAV.

## 3. Data

The proposed TAM is estimated using a sample dataset of the California Vehicle Survey conducted by California Energy Commission (2019). The sample dataset contains 3,723 individuals who are asked about their socio-demographic characteristics as well as their opinions about a number of attitudinal, perceptional, and preferential questions, which are discussed in the following sections.

### 3.1. Individuals' socio-demographic characteristics

The sample individuals are characterized by socio-economic attributes with the statistical distribution shown in Table 2. The survey respondents are 18 years of age and older pertaining to one of the three generations including Millennials (18 ≤ age < 34), Generation X (35 ≤ age < 64), and Baby Boomers (age ≥ 65). The ethnicity distribution of the dataset is so that 9.08% are Hispanic, Latino, or Spanish origin and 87.24% are not. Categorized based on the respondents' household annual income, almost one-fourth of the individuals are at low level (income < \$50K), almost half earn medium level (\$50k ≤ income < \$150K), and the rest belong to the high level (income ≥ \$150K). The respondents are further questioned about the highest education level, selecting from four different options ranging from no college to a professional degree.

### 3.2. Individuals' opinions on attitudinal, perceptional, and preferential questions

This subsection is structured in three parts including dependent variable, utilitarian and hedonic motivations, and attitude towards shared mobility.

*3.2.1. Dependent variable: Preference for owning AV over using SAV*

The individuals' response to the outcome variable is statistically distributed according to Figure 2. As observed, the majority are somewhat interested in either AV or SAV.

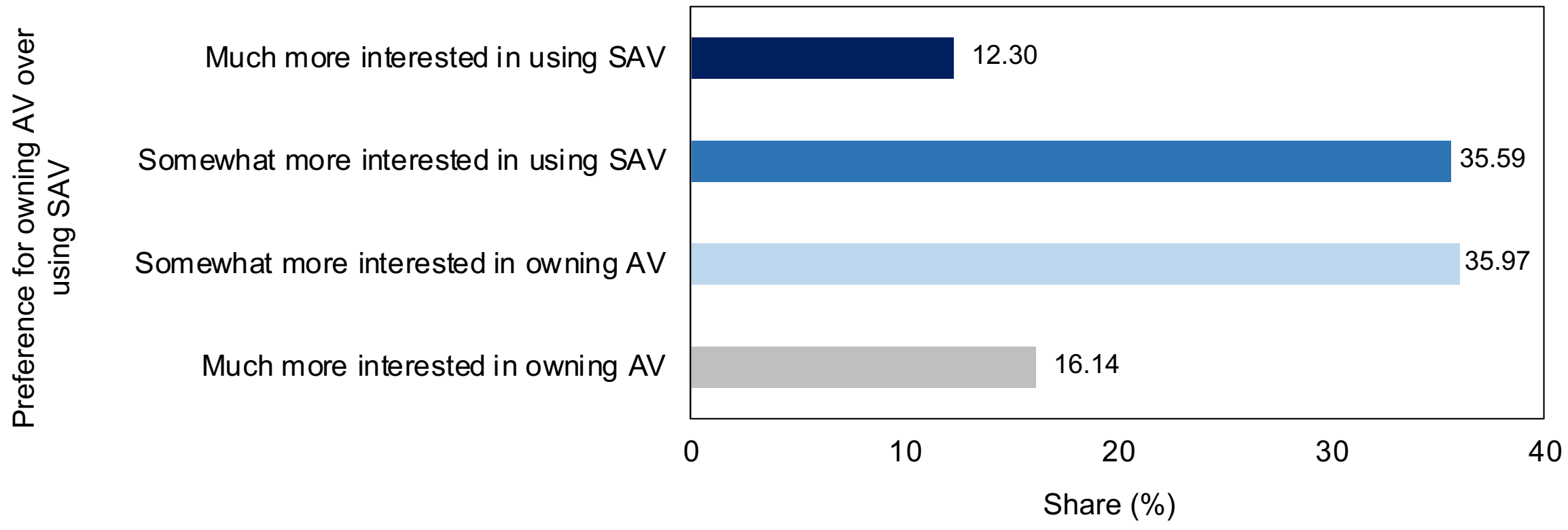

**Figure 2. Statistical distribution of the item on preference for owning AV over using SAV (#obs = 3,723)**

**Table 2: Sample data for individuals' socio-demographic characteristics (#obs = 3,723)**

| Variable | Category | #observations | Share (%) |
|---|---|---|---|
| Gender | Male | 1,962 | 53.30 |
| | Female | 1,719 | 46.70 |
| Generation | | | |
| | Millennials (18 ≤ age < 35) | 428 | 11.50 |
| | Generation X (35 ≤ age < 65) | 1,983 | 53.26 |
| | Baby Boomers (age ≥ 65) | 1,312 | 35.24 |
| Ethnicity | | | |
| | Hispanic, Latino, or Spanish origin | 338 | 9.08 |
| | Non-Hispanic | 3,248 | 87.24 |
| | Prefer not to answer | 137 | 3.68 |
| Race | White | 2,649 | 71.63 |
| | Asian | 543 | 14.68 |
| | African American | 126 | 3.40 |
| | American Indian or Alaska Native | 78 | 2.10 |

| Variable | Category | #observations | Share (%) |
|---|---|---|---|
| | Other | 302 | 8.16 |
| Employment status | | | |
| | Employed | 1,957 | 52.57 |
| | Unemployed | 1,467 | 39.40 |
| | Self-employed | 299 | 8.03 |
| Household annual income | | | |
| | Low income (income < $50K) | 699 | 20.56 |
| | Medium income ($50k ≤ income < $150K) | 1,782 | 52.43 |
| | High income (income ≥ $150k) | 918 | 27.01 |
| Educational attainment | | | |
| | High school graduate or less | 263 | 7.06 |
| | Technical school or some college | 1004 | 26.96 |
| | College graduate (A.S. or B.A.) | 1,127 | 30.27 |
| | Post graduate work or degree | 1,329 | 35.69 |

#### *3.2.2. Antecedent variables: Utilitarian and hedonic beliefs*

The statistical distribution of the individuals' responses to the questions on utilitarian and hedonic beliefs is shown in Figure 1. Each question presents a four-point Liker-scale response options to individuals which ranges from order 1 "strongly disagree" to order 4 "strongly agree." The individuals' responses to the questions are statistically distributed according to Figure 3. In response to four questions about PU of AVs, almost two-thirds either strongly or somewhat disagree with reducing time at their regular workplace to "work more in the self-driving car." The statement on "would send an empty self-driving car to pick up/drop off my child" received the most disagreement, especially strong than somewhat, with the share of 82.62%. The individuals' responses have almost an equal distribution over agreement and disagreement with two questions on traveling more often with AVs even when the person is tired, sleepy, or under the influence of alcohol/medications and seeing "a need for self-driving vehicles."

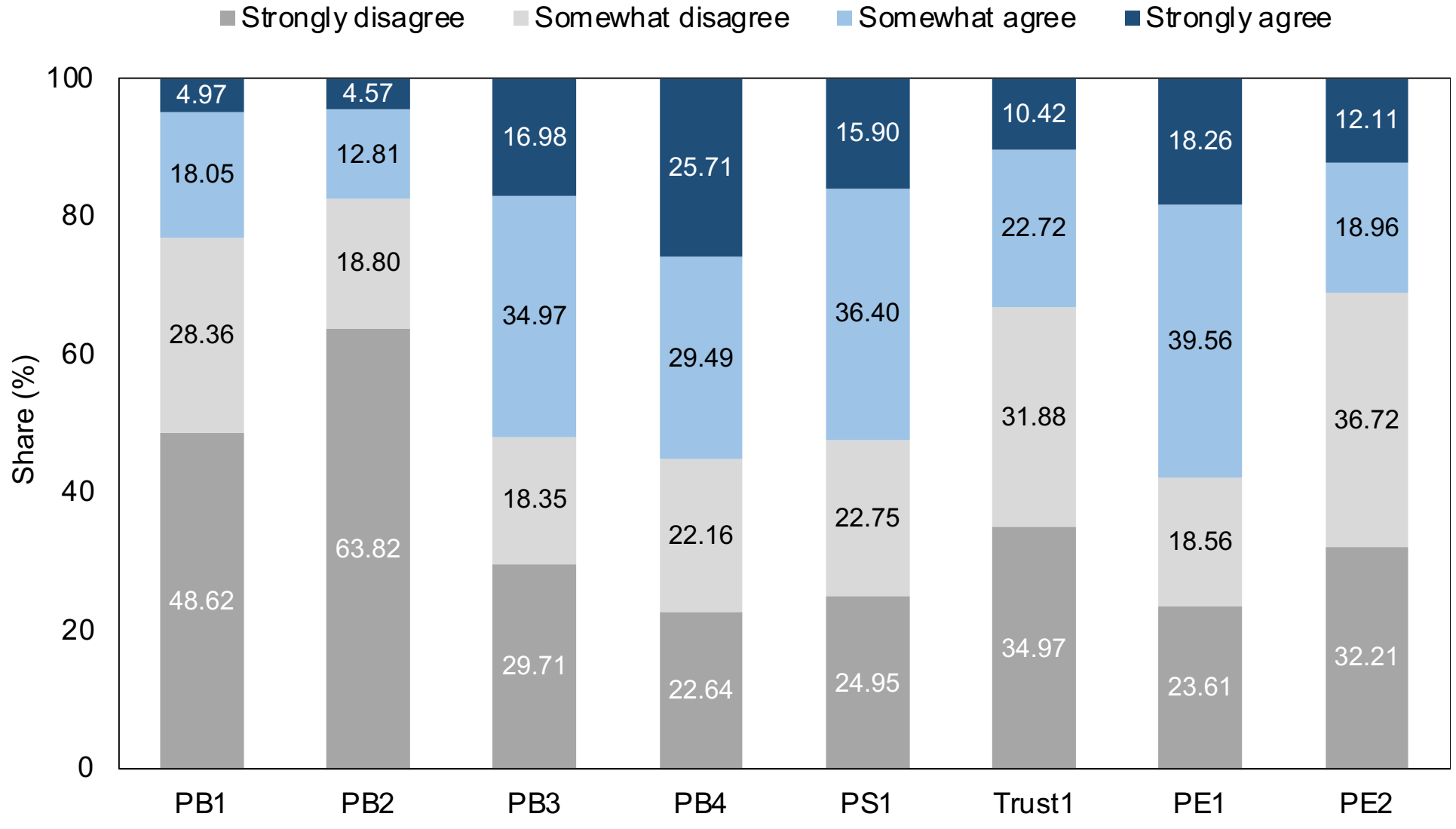

| Item | Definition |
|---|---|
| PB1 | I would reduce my time at the regular workplace and work more in the self-driving car. |
| PB2 | I would send an empty self-driving car to pick up/drop off my child. |
| PB3 | I would be able to travel more often even when I am tired, sleepy, or under the influence of alcohol/medications. |
| PB4 | I see a need for self-driving vehicles. |
| PS1 | I would accept longer travel times so the self-driving vehicle could drive at a speed low enough to prevent unsafe situations for pedestrians and bicyclists. |
| Trust1 | I would be likely to use shared driverless services (even at higher cost) because I would have no issues with sharing a vehicle with strangers. |
| PE1 | A self-driving vehicle would enable me to enjoy traveling more (e.g., watch scenery, rest). |
| PE2 | I would not miss the joy of driving and being in control. |

**Figure 3. Statistical distribution of items on utilitarian and hedonic beliefs (#obs = 3,723)**

The latent variable perceived safety is constructed on only one item which is the question on accepting to ride a long time in a self-driving vehicle at a low speed to prevent dangerous scenarios from occurring to pedestrians and bicyclists. The individuals' responses are observed to be almost equally distributed over strongly/somewhat agreement and strongly/somewhat disagreement. Similarly, the latent variable trust is built on only one-time capturing individuals' agreement with a statement on "I would be likely to use shared driverless services (even at higher cost) because I would have no issues with sharing a vehicle with strangers." The responses are observed mostly (66.85%) as disagreement either strongly or somewhat.

The latent construct explaining perceived enjoyment is built on two items. The sample distribution reveals that 57.82% of individuals agree with "A self-driving vehicle would enable me to enjoy traveling more (e.g., watch scenery, rest)." The second item asks individuals if they "would not miss the joy of driving and being in control." More than half of them (67.93%) indicate disagreement with the statement, meaning that they would miss the joy of driving and being in control.

### *3.2.3. Attitude towards shared mobility*

The survey collects information on the respondents' attitude towards shared mobility through measuring their experience with five car and ridesharing, two bikesharing, and three public transit modes in the residing area. For each mode, the respondents determine their experience using a four-point Liker-scale including order 1 "Not familiar with it", order 2 "Not Available where I live", order 3 "Available where I live, but never use it", and order 4 "Available where I live, and I use it." The statistical distribution of the individuals' responses is shown in Figure 4, where each angle represents one of the ten modes and the ordinal-level responses are depicted by four colored lines on the graph. Each marker on a line represents the percentage share of individuals in the sample dataset selecting the ordinal level associated with the line for the mode shown at the related angle.

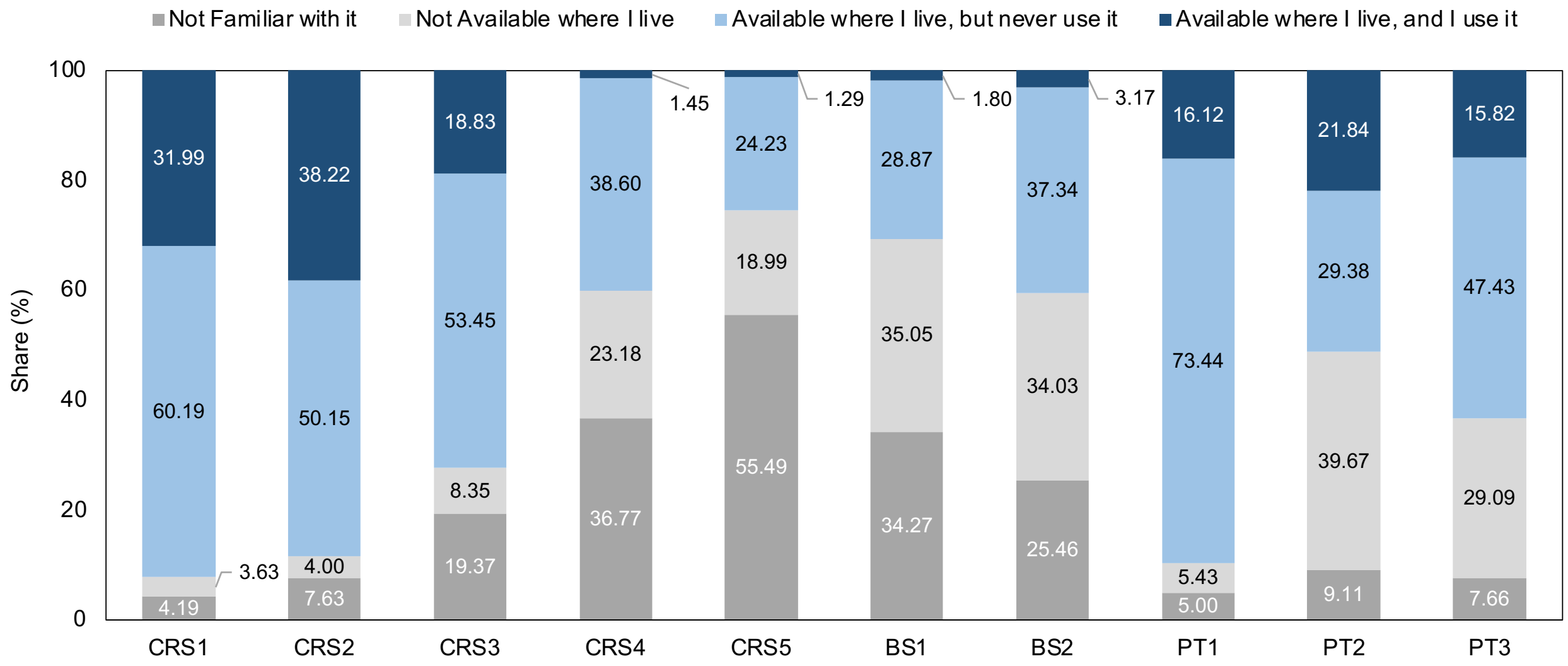

| Item | Definition |
|---|---|
| | Experience with ... for trips in the local area |
| CRS1 | Car rental |
| CRS2 | Ride hailing (e.g., Uber and Lyft) |
| CRS3 | Shared ride hailing (e.g., UberPool and LyftLine) |
| CRS4 | Car sharing (e.g., Car2Go and ZipCar) |
| CRS5 | Peer-to-peer car rental (e.g., GetAround and Turo) |
| BS1 | Bike sharing (e.g., Bay Area Bike Share) |
| BS2 | Shared eBikes or eScooters (e.g., Jump) |
| PT1 | Public bus |
| PT2 | Light rail/tram/subway (e.g., BART and LA Metro) |
| PT3 | Commuter train (e.g., Amtrak and Caltrain) |

**Figure 4. Statistical distribution of items on attitude towards shared mobility (#obs = 3,723)**

## 4. Results

### 4.1. Model evaluation

In reference to Hatcher and O’Rourke (2013) and Kline (1998), the model chi-square ratio is significant at 99.9% confidence interval. The goodness-of-fit index (GFI) and adjusted GFI meet the criterion compared to the threshold of higher than 0.9 (Cheung and Rensvold, 2002). The model surpasses the recommended value of Standardized root Mean Square Residual (SRMR) < 0.08 and Root Mean Square Error of Approximation (RMSEA) < 0.06 (Schreiber et al., 2006), with the indices of 0.17 and 0.13, respectively. However, it is reasonable since it does not favor models evaluating large sample sizes, with recommendations of reporting the matrix of correlation (Kline, 1998). Overall, the goodness-of-fit criteria suggest the model is a good fit to the dataset.

### 4.2. Model interpretation

#### *4.2.1. Estimated measurement equations*

The measurement equations connect each latent construct to the underlying items through the weight parameters. The estimation results are presented in Table 3 verifying the statistical significance of all parameters at 99.9% confidence interval. Of note is that the four items explaining perceived benefit have positive loadings with almost similar weights implying individuals’ acknowledgement of the four AV benefits. On the attitude towards shared mobility, the five items corresponding to car and ride sharing experience have significant factor loadings. The descending order of the highest estimated coefficients starts with shared ride-hailing, ride-hailing, carsharing, peer-to-peer car rental, and rental car.

Additionally, Table 3 presents the standard deviations (SD) of the scales used for assessing the construct reliability, composite reliability (CR), and the average variance extracted (AVE) to test the model’s convergent and discriminant validity through the factor loadings. To obtain the convergent and discriminant validity, CR and AVE need a recommended value higher than 0.60 and 0.40, respectively (Hatcher and O’Rourke, 2013). All constructs have an acceptable CR indicating a good reliability. Furthermore, all the AVE values meet the criteria except for one latent construct on car & ride sharing experience, though, meets the recommended values for the factor loadings and CR which signify the fair performance of the construct that in the Model (Hatcher and O’Rourke, 2013). Overall, these assessments confirm the convergent and discriminant validity of the measurement model.

#### *4.2.2. Estimated structural equations*

The structural equations yield the parameters showing the impact of the latent constructs on each other, as discussed in the section on the conceptual framework and the hypotheses. The estimation results are shown in Figure 5, depicting if the hypothesized relationships made in the study are supported. The significant and insignificant hypotheses are shown by solid and dashed arrows, respectively.

**Table 3. Estimation results of measurement equations**

| Latent construct | Factor loading | SD | CR | AVE |
|---|---|---|---|---|
| ***Dependent variable*** | | | | |
| Preference for owning AV over using SAV | | 0.946 | 0.984 | 0.989 |
| AV over SAV | 0.992*** | | | |
| ***Utilitarian beliefs*** | | | | |
| Perceived benefit (PB) | | 0.503 | 0.743 | 0.420 |
| PB1 | 0.629*** | | | |
| PB2 | 0.669*** | | | |
| PB3 | 0.685*** | | | |
| PB4 | 0.608*** | | | |
| Perceived safety of other road users than the AV passengers (PS) | | 1.01 | 0.990 | 0.991 |
| PS1 | 0.995*** | | | |
| Trust of strangers | | 0.942 | 0.951 | 0.952 |
| Trust1 | 0.975*** | | | |
| ***Hedonic beliefs*** | | | | |
| Perceived enjoyment (PE) | | 0.971 | 0.617 | 0.481 |
| PE1 | 0.899*** | | | |
| PE2 | 0.394*** | | | |
| ***Attitude towards shared mobility*** | | | | |
| Car and ride sharing experience (CRS) | | 0.540 | 0.682 | 0.307 |
| CRS1 | 0.372*** | | | |
| CRS2 | 0.603*** | | | |
| CRS3 | 0.651*** | | | |
| CRS4 | 0.601*** | | | |
| CRS5 | 0.500*** | | | |
| Bike sharing experience (BS) | | 0.535 | 0.706 | 0.546 |
| BS1 | 0.713*** | | | |
| BS2 | 0.764*** | | | |
| Public transit experience (PT) | | 0.312 | 0.670 | 0.405 |
| PT1 | 0.583*** | | | |
| PT2 | 0.693*** | | | |
| PT3 | 0.629*** | | | |

**Note: *** significant at 99.9% confidence interval.**

Among utilitarian beliefs, perceived benefit is found to have a positive role in the preference for AV over SAV which is also confirmed by the previous studies thoroughly reviewed in Keszey (2020). In contrast, trust of passengers is negatively associated with the outcome construct meaning that those who trust in sharing a vehicle with strangers do not prefer AV over SAV. However, the absolute value of the former factor is larger than the latter showing the associated higher impact which is in line with the prior studies having the perceived benefit factor as a main factor contributing to propensity towards AV (Hegner et al., 2019; Panagiotopoulos and Dimitrakopoulos, 2018). The results further reveal the insignificant role of the perceived safety for other road users than the AV passengers. Moreover, the only relationship between the utilitarian constructs is found significant on the role of perceived benefit in trust showing a strong association.

Furthermore, the hedonic beliefs reflected in perceived enjoyment has a significant and the highest impact on the outcome construct, which verifies findings of the relevant studies on AV acceptance and adoption (Madigan et al., 2017). Among the hypothesis made on attitudes towards shared mobility, experience with car and ride sharing and public transit are detected with negative impacts meaning that those who have experience with these services less prefer AV over SAV. This indicates the positive role of these services on encouraging their users to use SAVs in the future. The bikesharing experience latent constructs is found insignificant in the model, which might be due to the overall negative attitude of bikesharing users towards non-active travel modes.

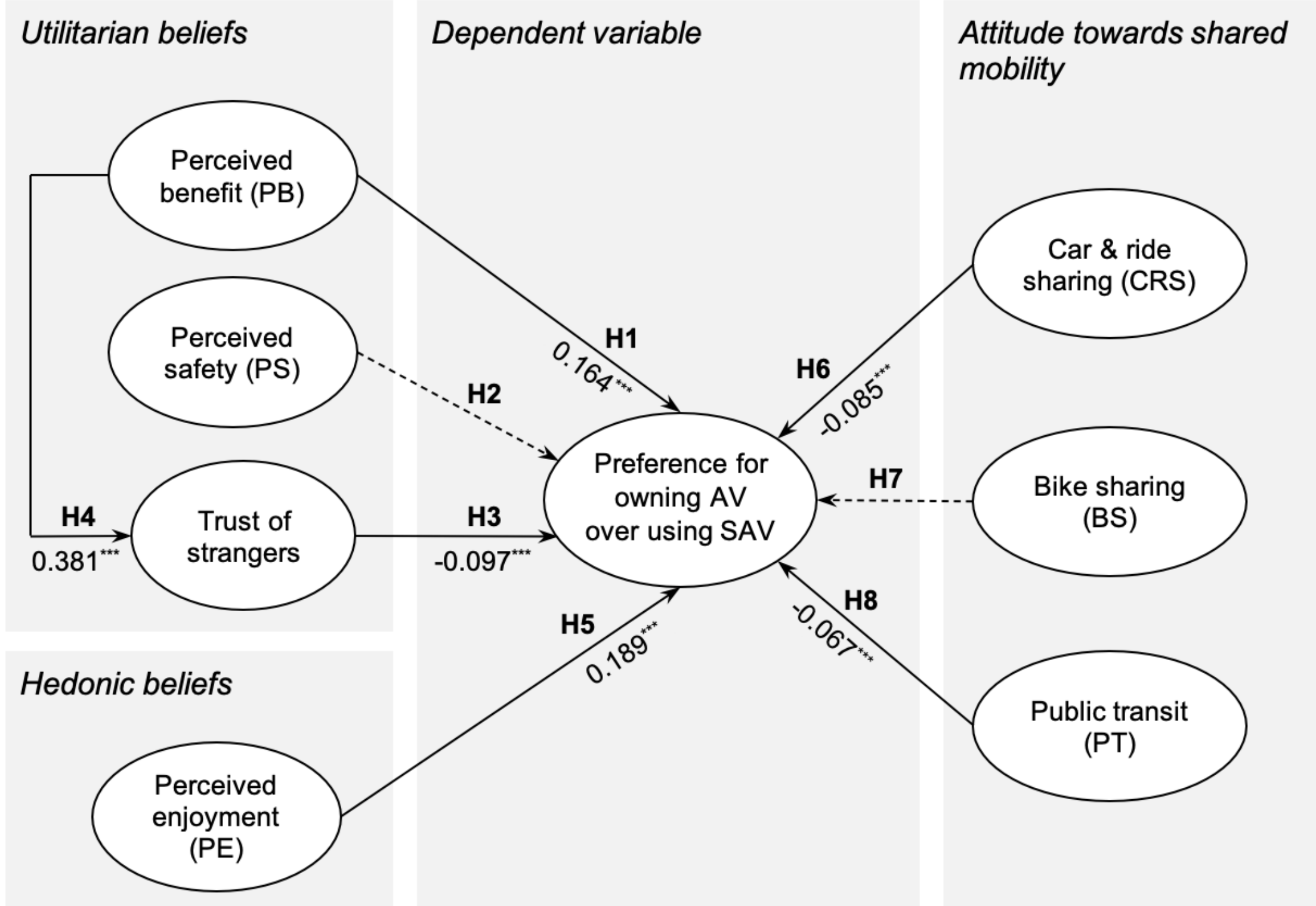

**Note: *** significant at 99.9% confidence interval. Dashed lines indicate insignificant effect.**

**Figure 5. Estimation results of the proposed model**

## Conclusions

The emerging automated vehicle (AV) is a technological innovation with the potential of transforming transportation systems promising various benefits to the society, especially if used in shared systems (SAV). However, there are classes of individuals intend to own an AV as a private asset than using SAV. To have a better understanding of the behavior of those classes, there is a research need to look into the behavioral tendency towards owing versus using shared AV with a psychological perspective to recognize the influential factors. In order to do so, the present study proposes a modeling framework built on *behavioral psychology* which explains individuals' preference for AV over SAV by their utilitarian beliefs, hedonic beliefs, and attitude towards shared mobility. By estimating the proposed model on a sample dataset collected in the State of California, the results reveal that preference for AV over SAV is affected in a descending order by hedonic beliefs, reflected in perceived enjoyment, utilitarian beliefs, particularly perceived beliefs and trust of stranger, and attitude towards shared mobility, especially car and ridesharing experience.

## Acknowledgments

This work was funded by the US National Science Foundation (Award No. 2112650). The authors are solely responsible for the findings of this research. The opinions expressed are solely those of the authors, and do not necessarily represent those of the US National Science Foundation. The authors would like to acknowledge the USDOW Dwight D. Eisenhower Transportation Fellowship Program for the student support provided and are also grateful to California Energy Commission for providing the study dataset.

## Author Contributions

The authors confirm contribution to the paper as follows: study conception and design: F. Nazari, Y. Soto, M. Noruzoliaee; analysis and interpretation of results: F. Nazari, Y. Soto, M. Noruzoliaee; draft manuscript preparation: F. Nazari, Y. Soto, M. Noruzoliaee. All authors reviewed the results and approved the final version of the manuscript.